# SIP Server Load Balancing Based on SDN

Ahmadreza Montazerolghaem

Department of Computer Engineering,
Quchan University of Technology,
Quchan, Khorasan Razavi, Iran
Ar.montazer@qiet.ac.ir

*Abstract*— Session Initiation Protocol (SIP) grows for VoIP applications, and faces challenges including security and overload. On the other hand, the new concept of Software-defined Networking (SDN) has made great changes in the networked world. SDN is the idea of separating the control plane from the network infrastructure that can bring several benefits. We used this idea to provide a new architecture for SIP networks. Moreover, for the load distribution challenge in these networks, a framework based on SDN was offered, in which the load balancing and network management can be easily done by a central controller considering the network status. Unlike the traditional methods, in this framework, there is no need to change the infrastructures like SIP servers or SIP load balancer to implement the distribution method. Also, several types of load distribution algorithms can be performed as software in the controller. We were able to achieve the desired results through simulating the three methods based on the proposed framework in Mininet.

I. INTRODUCTION

Session Initiation Protocol (SIP) [1] is a signaling protocol to handle a variety of applications, including Voice over IP (VoIP) and Instant Messaging (IM) [2] for calls establishment and termination. Two important components in this protocol are SIP user agents and servers. In SIP, call requests from user agents are sent to the servers. Since SIP server capacity is limited, the large number of the SIP user agents could possibility cause the servers overload. One way to deal with this phenomenon is the distribution of the user requests among the servers. Therefore, as Figure 1 illustrates, a load balancer among the user agents and servers is required. The load balancer must be informed about the available capacity of each server as a real time and determines the best server to service the request on its basis. A network of switches is responsible for sending the requests to the load balancer. Figure 1 shows that, there is a possibility of the bottleneck in the load balancer because all the requests pass through it. Given the challenge, we present a new framework for SIP networks to distribute the load.

The rest of this article is as follows: In Section II, we have a quick overview of the previous literature. In Section III, we introduce SDN network. The proposed framework is presented in Section IV and in Section V, we evaluate its performance. The conclusion of this article is done in section VI.

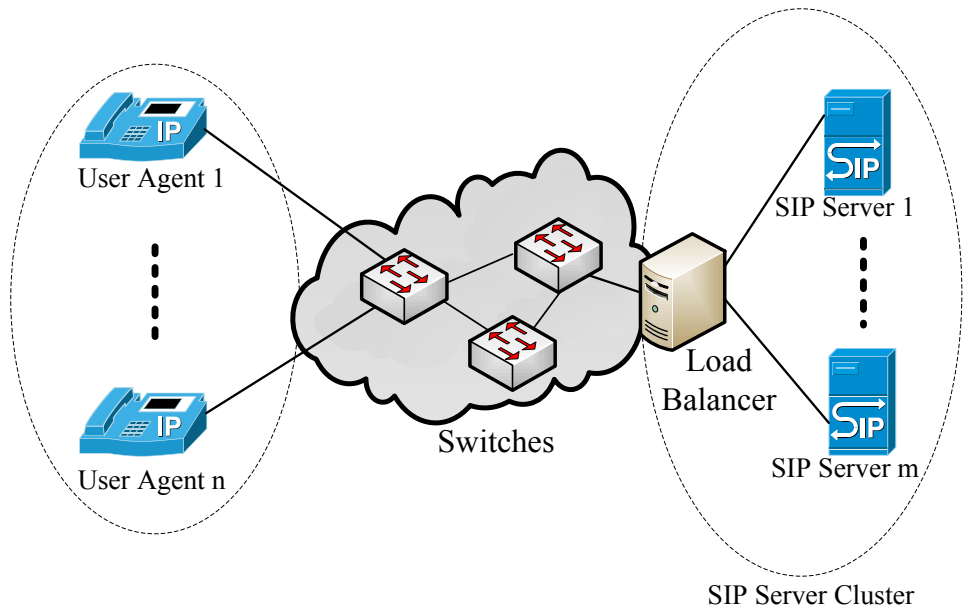

Fig. 1. Traditional architecture of load distribution in SIP

## II. RELATED WORKS

In [3], a load balancing for SIP is provided. In this paper, based on the recipient of the call, requests are routed to servers. To allocate calls to the servers, a hash function is used. A key problem in this approach is that fairness is difficult. Also, this approach does not adapt well to the call changes. Article [3] explains the high availability and proposes how to fix the failures.

A number of commercial products support SIP load balancing and one of them is "Nortel Networks' Layer 2–7 Gigabit Ethernet Switch Module for IBM Blade Center" [4]. Basic information of this product is available, but what the load-balancing algorithm uses is not clear.

Considerable research has been done in the field of load balancing for HTTP requests [5]. One of the most original articles in the field describes the scaling manner of NCSA's website using DNS [6]. In [7], the advantages of using an explicit load balancing of round robin DNS is shown. Load balancer in that article is not informed about the request content because the content does not examine any request. In papers [8], [9] and [10], a load balancer with the knowledge of the content is discussed. Such a balancing reviews the request itself to decide on the route. In [11] and [12], the load balancing in the websites having high access rate in reality has been addressed. In [13] and [14], client-side algorithms for assign the requests to a server have been provided. In [15], a load balancing for cluster of web servers is presented based on the requests size. Least work left and joining the shortest queue to assign the tasks to the servers are also provided in [16] and [17] and [18]. However, these articles do not show how a load balancer can estimate the least work left for a SIP server as reliable mechanism.

The architecture of all the articles in this field is in accordance with Figure 1. This means that to distribute the load between the servers in SIP, an entity called the load balancer is always used. The difference between the various articles is in the work of this entity and its algorithm. In this article, inspired by the concept of SDN, a new framework for the distribution of the load between the SIP servers is offered. To our knowledge, this is the first research on SIP overload control through SDN approach.

## III. SOFTWARE DEFINED NETWORK TECHNOLOGY

SDN is new network architecture [19]. Figure 2 shows the basic structure of SDN. As can be seen in Figure 2, the whole network is made of two main parts: the control plane and the data plane.

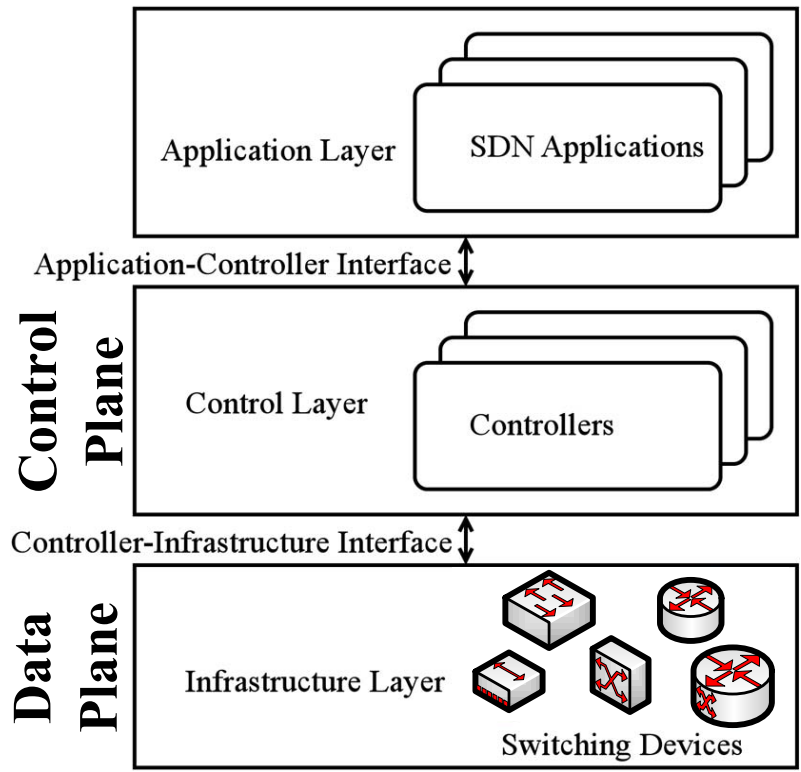

Fig. 2. SDN Architecture

All controls are on the control plane and it is in fact the brain of the SDN. The control plane makes possible the network management for the user software in the application layer. The main policies can be made in the control plane. In the data plane, there are the main hardware and software elements, such as switches, routers and firewalls which act like a traditional network. The connections between these elements are through common media such as fiber optic or copper cable and each of which has its own performance. One of the advantages of SDN is that the switches and routers are not tied to a physical location or a particular brand. The connections between the different layers in SDN architecture are established through open and standard interfaces (including OpenFlow) [20-22]. Thus, there are two main components in an architectural SDN:

- Forwarding elements (SDN switches)
- SDN controllers

A forwarding device is a hardware or software that is specifically charged with the task of forwarding the packages based on the flow table. The flow table contains rule, action and counter. The controller like a network operating system controls the forwarding devices and facilitates the automated management of the networks. In other words, the rules that the switches must follow are provided by the controller. These rules depend on the policies of the application software in the application layer. The action field determines the behavior of the switches with the packets matched with the rule specified. The counter is also used for counting these packets [23-25].

Network management, economic efficiency and adaptability are the architectural features of SDN. SDN also makes possible the configuration of the network devices from a central point and automatically through the software. In this way, the entire network can be programmatically and dynamically configured based on the network status [21].

## IV. SIP Server Load Balancing Based on SDN

In this section, we first introduce the proposed architecture for SIP networks. This architecture is shown in Figure 3. As it is clear in this figure, by replacing the traditional switches of the network with SDN switches, its advantages can be used in SIP networks. In other words, through the central management of SDN switches by a controller, the policies (such as policies of "routing", "traffic engineering" and "security") can be performed without changing the network infrastructure (such as SIP servers and user equipment). Previously, it was necessary to change the network infrastructure devices to implement each of these policies in SIP network. For example, to apply an overload control policy, the configuration of each of the SIP servers should have been changed and made the management and maintenance of the network difficult and made the network inflexible. Using the proposed architecture in the present article, the network policies can be easily implemented on the central controller as software and installed on the switches as rules using the OpenFlow instructions. Therefore, all SIP network controls are done without changing the servers in the controller.

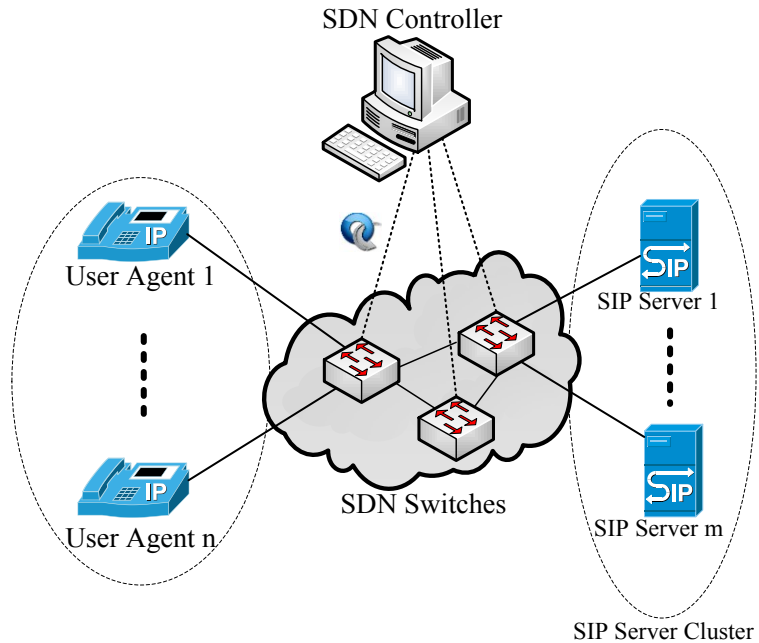

Fig. 3. Proposed Architecture

This architecture is applicable for all the challenges in SIP network (such as security, overload, load distribution, etc.) but to prepare the preparations for the development of this architecture in the future, the rest of this article will be allocated to the load distribution challenges. Figure 4 shows the proposed framework for SIP load balancing according to Figure 3. The proposed controller in this context includes the following components:

- **SDN applications (e.g., load balancing):** It includes load distribution algorithm. The input is the information regarding network manager and server manager.
- **Network manager:** It provides a global view of network topology.
- **Server manager:** It monitors the server loads using the counter field in the flow table switches.
- **Flow manager:** It manages and routes the flows toward the best server by setting the appropriate rules according to the load balancing application.

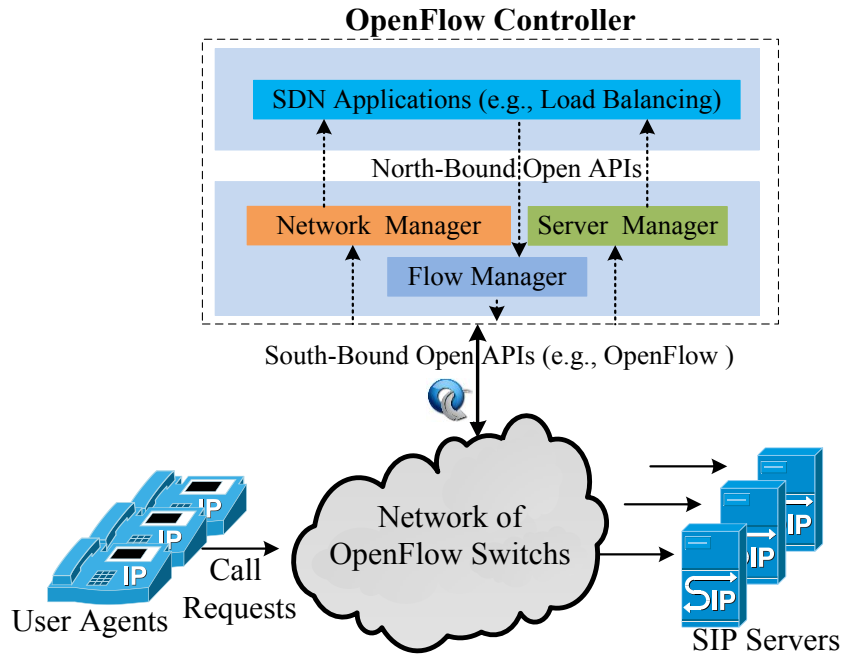

Fig. 4. Proposed framework for load distribution

In the following, three methods for used in the load balancing applications are explained, and then in section V, the proposed framework is evaluated using these three methods.

***The first method - Random*:** A server is chosen randomly and a new request is sent to the server.

***The second method - Round robin*:** The servers are attributed to the requests with no priority and only through their rotation.

***The third method - Least request*:** The new request is sent to the server with the least load. To identify the server with the lowest load, the counter field can be used.

## V. SIMULATION AND PERFORMANCE EVALUATION

Mininet emulator was used to simulate the proposed framework in Linux. The studied topology in Mininet is shown in Figure 5. This topology includes a floodlight controller, a switch, three servers with specific IP address ($\overline{\text{Host}}$) and *n* user agents (Host). Moreover, the experiments have been done on an Inspiron1525 DELL laptop with an Intel Core 2 Duo CPU and a 3 GB RAM, a 32-bit Windows seven operating system and Oracle VirtualBox as virtual machine.

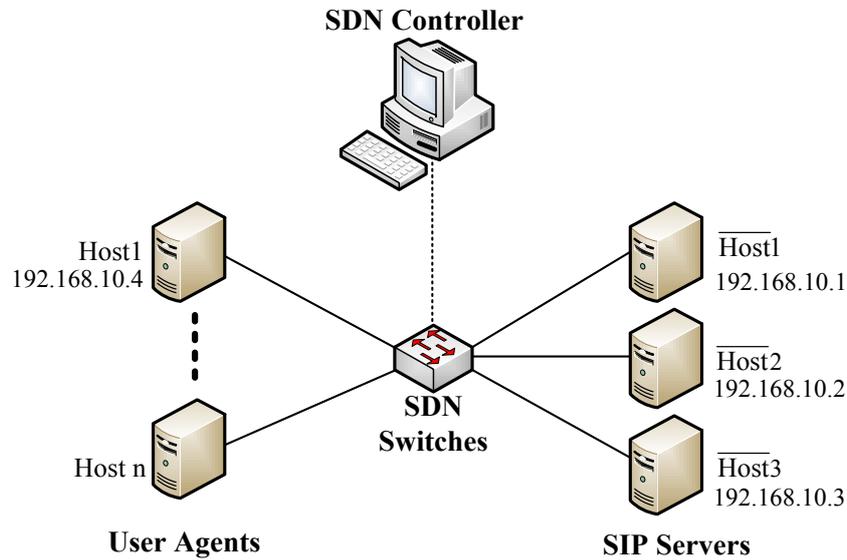

Fig. 5. Test topology in Mininet

According to Table I, three scenarios have been designed to evaluate the proposed framework. This table shows the number of the background requests in each server per scenario. In addition, 200 requests have been created by the user agents for about 50 seconds.

TABLE I. NUMBER OF BACKGROUND REQUESTS

|  | Server 1 | Server 2 | Server 3 |
| --- | --- | --- | --- |
| Scenario 1 | 50 | 50 | 50 |
| Scenario 2 | 100 | 50 | 25 |
| Scenario 3 | 200 | 50 | 0 |

Two criteria have been used for the assessment: "average response time" and "throughput". Average response time is the time between sending a request and receiving a response. Throughput is the number of the responded requests in the time unit.

Figures 6, 8 and 10 show the average response time and Figures 7, 9 and 11 show the throughput of the methods in the three scenarios. As you can see, the Least Request method has a less average response time and a greater throughput than the other two methods. However, in Scenario 1, the three methods are close to each other with equal number of background requests. In Scenarios 2 and 3 because of an unequal number of the background requests, more differences in the methods can be observed. For example, in Scenario 3, since the background load of Server 1 is greater than the two other servers, the new request should not be sent to Server 1. This is while the Random and Round robin methods do not consider such an important point. In this scenario, the Least Request method chooses Server 3 for the new requests.

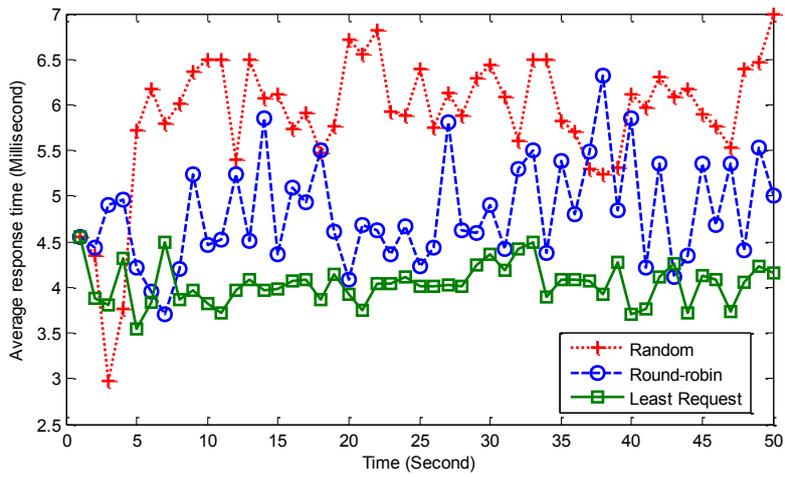

Fig. 6.  Average response time in Scenario 1

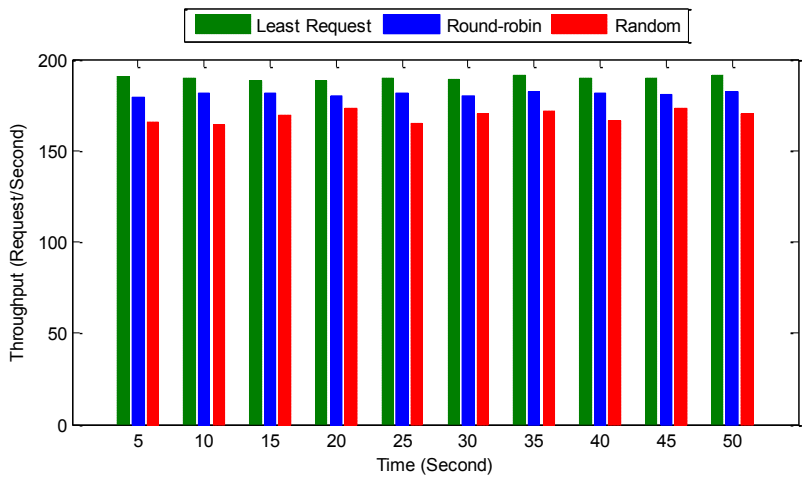

Fig. 7.  Throughput in Scenario 1

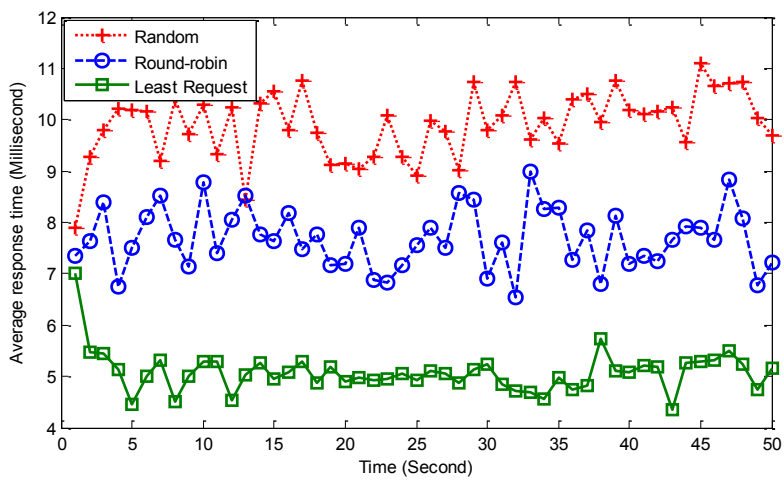

Fig. 8.  Average response time in Scenario 2

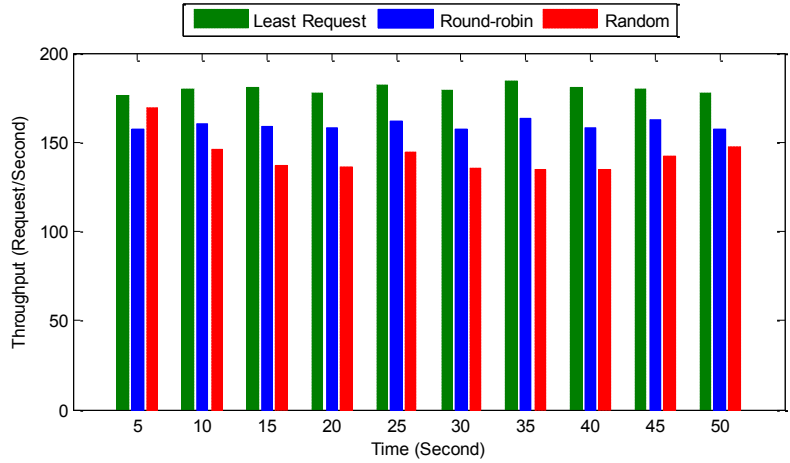

Fig. 9. Throughput in Scenario 2

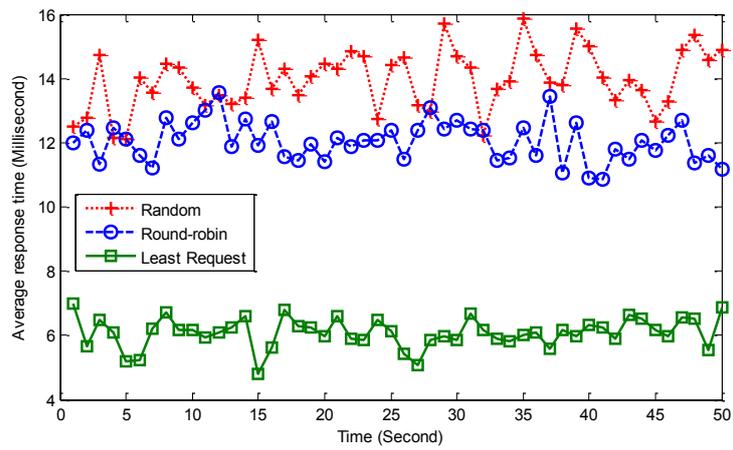

Fig. 10. Average response time in Scenario 3

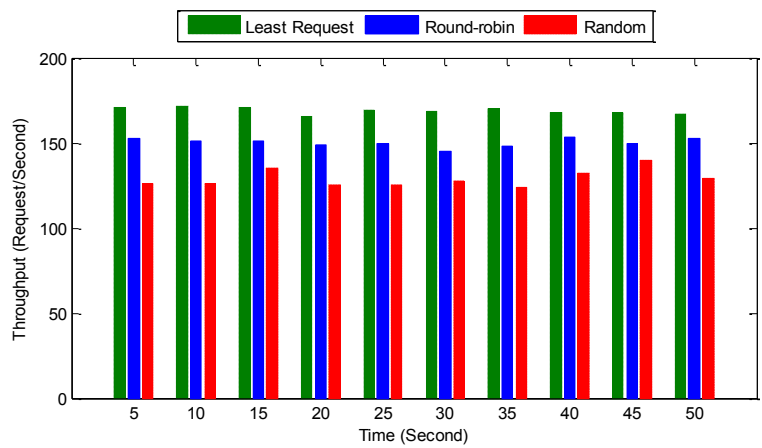

Fig. 11. Throughput in Scenario 3

In Figures 12 and 13, the effect of the increase of *n* (user agents) is shown. As expected, by increasing *n*, the number of the requests increases and therefore the average response time increases while the throughput decreases. But these changes are less in the Least Request method. This means that this method is able to take advantage of the information of the switches to have a proper load distribution.

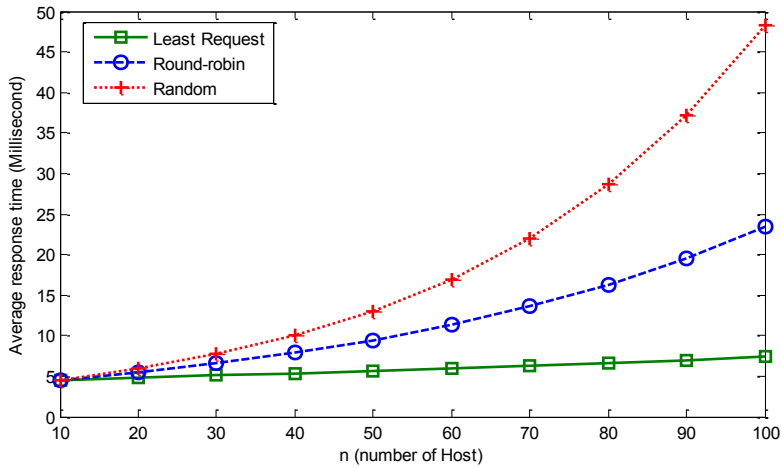

Fig. 12. Average response time by increasing *n*

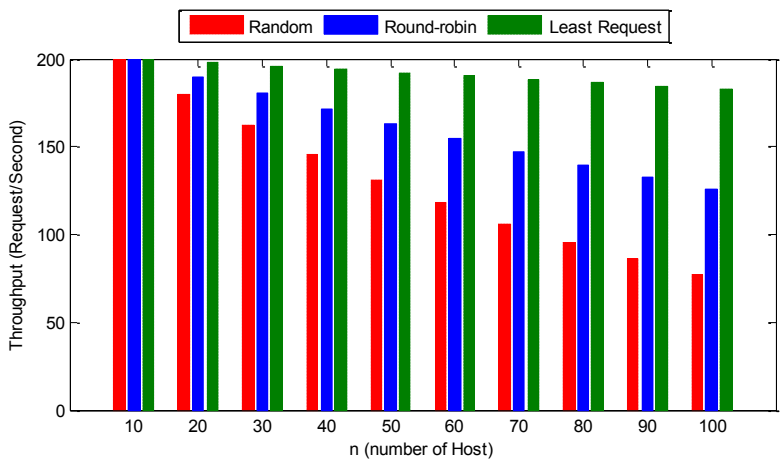

Fig. 13. Throughput by increasing *n*

## VI. CONCLUSIONS AND FUTURE WORK

SIP networks has challenges, including security, and overload which are continuously researched. On the other hand, the concept of SDN has recently made great changes in the network. SDN is the idea of separating the control plane from the data plane which has many benefits. We have used this idea to provide a new architecture for SIP networks. We have also provided a framework based on SDN for the challenge of SIP load balancing in the network. The important point is that in the traditional methods, the load balancer could only use one algorithm while in our proposed framework, the load distribution method can be changed regarding the network status and the network management can be done from a central point. Based on the proposed framework, several types of load distribution algorithms can be performed as software in the controller without changing the network infrastructure and their effects can be investigated. In other words, researchers in the field of SIP load distribution can code the algorithms in the controller without changing the data plane. We have achieved the desired results through simulating the three methods based on the proposed framework. However, the practical implementation of this framework may be challenging and will be the focus of our future work.